\documentstyle{aipproc}

\newcommand{\hepex}[1]{(hep-ex/#1)}
\newcommand{\hepph}[1]{(hep-ph/#1)}
\def\prl#1#2#3{\frenchspacing{\it Phys. Rev. Lett. }{\bf #1}, #2 (19#3)}
\def\pr#1#2#3{\frenchspacing{\it Phys. Rev. D}{\bf #1}, #2 (19#3)}

\def\prep#1#2#3{\frenchspacing{\it Phys. Rep. }{\bf #1}, #2 (19#3)}
\def\arnps#1#2#3{\frenchspacing{\it Ann. Rev. Nucl. Part. Sci. }{\bf #1}, #2 (19#3)}

\newcommand{\etal}{{\em et al.}}
\newcommand{\ie}{{\em i.e.}}

\newcommand{\gevcc}{\hbox{ GeV}\!/\!c^2}
\newcommand{\gev}{\hbox{ GeV}}
\newcommand{\gevc}{\hbox{ GeV}\!/\!c}

\newcommand{\mev}{\hbox{ MeV}}
\newcommand{\mevc}{\hbox{ MeV}\!/\!c}
\newcommand{\mevcc}{\hbox{ MeV}\!/\!c^2}
\newcommand{\tev}{\hbox{ TeV}}

\newcommand{\cm}{\hbox{ cm}}

\newcommand{\fb}{\hbox{ fb}}
\newcommand{\flux}{\hbox{ [cm}^{-2}\hbox{s}^{-1}]}
\newcommand{\lum}{\hbox{ cm}^{-2}\hbox{s}^{-1}}
\def\ltap{\mathop{\raisebox{-.4ex}{\rlap{$\sim$}} 
\raisebox{.4ex}{$<$}}}

\newcommand{\cfrac}[2]{\textstyle \frac{#1}{#2}}
\newcommand{\mumu}{$\mu^{+}\mu^{-}$ collider}

\begin{document}
\begin{flushright}
\rule{0pt}{36pt} FERMILAB--CONF--98/073--T 
\vspace{-30pt}
\end{flushright}
\title{Physics with a Millimole of Muons}

\author{Chris Quigg}
\address{Fermi National Accelerator Laboratory\thanks{Fermilab is 
operated by Universities Research Association Inc.\ under Contract 
No.\ DE-AC02-76CH03000 with the United States Department of Energy.}\\
P.O. Box 500, Batavia, Illinois 60510 USA \\ E-mail: \textsf{quigg@fnal.gov}}

\maketitle

\begin{abstract}
The eventual prospect of muon colliders reaching several TeV encourages us 
to consider the experimental opportunities presented by 
very copious stores of muons, approaching $10^{21}$ per year. I 
summarize and comment upon some highlights of the \textit{Fermilab Workshop 
on Physics at 
the First Muon Collider and at the Front End of a Muon Collider.}  
Topics include various varieties of $\mu\mu$ colliders, $\mu p$ 
colliders, and applications of the intense neutrino beams that can be 
generated in muon storage rings.
\end{abstract}

\section*{Introduction}
The initial appeal of a \mumu\ is that it may provide a possible path 
to a few-TeV lepton-lepton collider to address the great issue of our 
age, the character of the mechanism that breaks electroweak symmetry.  
It is a commonplace that lepton colliders and hadron colliders offer 
complementary means to explore the nature of electroweak symmetry 
breaking \cite{pm,je}.  It is widely agreed that the rise of 
synchrotron radiation causes circular electron machines to become 
impractical for energies above a few hundred GeV.  Linear colliders 
are therefore under development for c.m.\ energies from a few hundred 
GeV to about $1.5\tev$.  I think it possible that linear-collider 
technology may only be interesting for about one decade in energy; the 
growth path beyond $1\hbox{ to }2\tev$ is not clear.  But it is a very 
interesting decade in energy, over which we expect to learn the secrets of 
electroweak symmetry breaking.  That is why there is such intense 
interest in the linear-collider approach.  In contrast, the 
extrapolation of a \mumu\ to several TeV per beam seems 
straightforward---if a \mumu\ can be made to work at all 
\cite{snowmass,status}.  If the small size of a \mumu\ is an 
indication of its cost, which is by no means established, a \mumu\ 
might even offer a less costly way to a modest-energy lepton collider.  
Taken together, these two possibilities offer a strong incentive to 
investigate the technology of a \mumu.

Once the technological possibility of a muon collider is raised, 
there are many interesting possibilities to contemplate \cite{bbgh}.  On the way 
to the ultimate prize of a 2--4-TeV collider, we may consider a 
high-luminosity $Z$ factory and machines to operate near the 
$W^{+}W^{-}$ and $t\bar{t}$ thresholds, as well as a machine with $\sqrt{s} 
\approx \cfrac{1}{2}\tev$ to explore details of a supersymmetric or 
technicolor world for which the first indications have been found 
elsewhere.  A \mumu\ also offers the unique possibility of a Higgs 
factory where detailed measurements not possible elsewhere could be 
undertaken.  The front end of a muon collider offers a host of 
possibilities of its own, including intense low-energy 
hadron beams, a copious source of low-energy muons, and the neutrino 
beams of unprecedented intensity and unusual flavor composition 
that emanate from stored muons.  A muon collider in the neighborhood 
of a hadron storage ring opens the possibility of high-luminosity $\mu 
p$ collisions as well.

Many of these possibilities have been explored at this Workshop, 
which I found notable for the fact that the participants actually did 
some original work.  My first---and most important---conclusion to the 
Workshop is that there are many \textit{interesting} physics topics to think 
about.

\subsection*{The Case for Muons}
The muon is massive: $m_{\mu}\approx 106\mevcc \approx 207 m_{e}$.  
Compared to electrons in a circular machine of given radius, muons of 
the same energy lose far less energy to synchrotron radiation, by a 
factor $(m_{e}/m_{\mu})^{4}\approx 5.5 \times 10^{-10}$.  A 
crippling problem for electron machines---and the reason we turn to 
linear colliders---is of negligible importance for a muon machine.

In common with the electron, the muon is an elementary lepton at our 
current limits of resolution.  Its energy is not shared among many 
partons, so the muon is a more efficient delivery vehicle for high 
energies than is the composite proton.  

Because the muon is massive, and can be accelerated efficiently in 
circular machines, and because we can probe the 1-TeV scale with muons 
of a few TeV, as opposed to protons of several tens of TeV, a muon 
collider can be small.  If a muon collider proves technically feasible, 
we need to discover whether small translates to inexpensive---both in 
absolute terms and compared to other paths we might take to high 
energies.

Beyond the suggestion of these practical advantages, muons offer a 
possibly decisive physics advantage.  The great seduction of a First 
Muon Collider is that the cross section for the reaction
$\mu^{+}\mu^{-} \rightarrow H$, 
direct-channel formation of the Higgs boson, is larger than the cross 
section for $e^{+}e^{-}\rightarrow H$ by a factor 
$(m_{\mu}/m_{e})^{2} \approx 42,750$.  This is a very large factor.  
The tantalizing question is whether it is large enough to make 
possible a ``Higgs factory'' with the luminosities that may be 
achieved in \mumu s.  In $e^{+}e^{-}$ collisions, of course, the 
$s$-channel formation cross section is hopelessly small.  That is why 
the associated-production reaction $e^{+}e^{-}\rightarrow HZ$ has 
become the preferred search mode at LEP--2.

The properties of the muon also raise challenges to the construction 
and exploitation of a \mumu.  The muon is not free: it doesn't come 
out of a bottle like the proton or boil off a metal plate like the 
electron.  On the other hand, it is readily produced in the decay 
$\pi \rightarrow \mu\nu$.  Still, gathering large numbers of muons in 
a dense beam is a formidable engineering challenge, and the focus of 
much of the R\&D effort over the next few years.  The muon is also 
not stable.  It decays with a lifetime of 2.2~$\mu$s into $\mu^{-} 
\rightarrow e^{-}\bar{\nu}_{e}\nu_{\mu}$.  We must act fast to 
capture, cool, accelerate, and use muons, and must be able to 
replenish the supply quickly.  Multiply 2.2~$\mu$s by whatever 
Lorentz $(\gamma)$ factor you like for a muon collider, it is still a 
very short time.

The muon's decay products complicate experimentation as well.  Just to 
indicate the dimensions of the problem, in a $2\oplus 2\,$-TeV collider with 
$2\times 10^{12}$ muons/bunch, every meter the bunch travels sees 
$2\times 10^{5}$ decays, with an average electron energy of about 
$700\gev$.

Finally, the neutrinos emitted in $\mu$ decay may constitute a 
radiation hazard.  You need not fear the neutrinos themselves.  The 
interaction length of a 100-GeV neutrino is about 25 million 
kilometers in water, so it has only about 1 chance in $10^{11}$ of 
interacting in the column depth of your body.  The potential hazard 
comes from neutrino interactions in the Earth surrounding a \mumu, 
which generate hadronic showers.  Estimates suggest that the potential 
radiation dose from these showers becomes a serious concern for 
$E_{\mu} \approx 1\hbox{ - }2\tev$.

\subsection*{The Big Questions for \mumu s}
When we discuss whether there should be muon colliders in our future, 
we must answer a number of important questions.

\begin{itemize}
	\item  What machines are possible? When? At what cost?

	\item  What are the physics opportunities?

	\item  Can we do physics in the environment? 
	(What does it take?)

	\item  How will these experiments add to existing knowledge not just 
	in the abstract, but \textit{when they are done?}
\end{itemize}
These questions are not the unique concern of a muon collider, but 
need to be addressed for any new accelerator we might contemplate.  I 
would like to underscore the importance of the last question: it is 
crucially important to try to judge what will be known from ongoing 
experiments and initiatives already launched at the moment that a new 
experimental tool could be ready.  What seems like essential 
information---if we could have it today---may fade in significance a 
decade or more hence.  Our goal must be to develop the means to do 
experiments that can change the way we think.  It is worth keeping in 
mind Bob Palmer's estimate that a First Muon Collider might be in 
operation around the year 2010 \cite{rbp}.
  
\subsection*{The Focus of This Workshop}
The Workshop on Physics at the First Muon Collider and at the Front 
End of a Muon Collider was organized around nine working groups.  One 
dealt with accelerator issues, concentrating on the design of a proton 
driver for the Fermilab site.  Progress on an RF system, longitudinal 
space-charge effects, the formation of short bunches a few ns in 
length, and instability questions was reported by Bob Noble 
\cite{noble}.  Four working groups addressed physics prospects for 
muon colliders.  They were organized around Higgs and $Z$ factories 
\cite{higgz}, top physics \cite{tops}, supersymmetry \cite{susy}, and 
strong dynamics \cite{techni}.  Four more working groups explored the 
physics interest of beams associated with the front end of a muon 
collider.  Those groups considered low-energy hadron physics 
\cite{hadrons}, neutrino physics \cite{neutrino}, deep inelastic 
scattering \cite{dis}, and low-energy muon physics \cite{lemur}.

\subsection*{The Front End of a Muon Collider}
The Front End of a Muon Collider consists of four basic elements.
\begin{itemize}
	\item A high-intensity proton source.  An example design developed for 
	the Fermilab site ends in a rapid-cycling synchrotron that delivers 16-GeV 
	protons at 15~Hz \cite{driver}.  In each cycle, two bunches of $5 
	\times 10^{13}$ protons are accelerated, for a total of $1.5 \times 
	10^{22}$ protons per year.  That is about $10^{3}$ the number of 
	protons delivered at $8\gev$ by the Fermilab Booster.

	\item A system for pion production, collection, and decay.  Charged 
	pions created in the collision of the proton beam with a target are 
	confined in a high-field solenoid and guided into a 20-meter-long decay channel 
	within a 7-Tesla solenoid that keeps the muons from escaping.  Such a 
	system might yield about 0.2 $\mu^{+}$ and $\mu^{-}$ per proton, or 
	about $10^{13}$ $\mu^{+}$ and $\mu^{-}$ per cycle, for a total of 
	about $1.5 \times 10^{21}$ $\mu^{+}$ and $\mu^{-}$ per year.

	\item A muon cooling channel to concentrate the muons in 
	six-dimensional phase space.  It is hoped that an ``ionization 
	cooling'' system \cite{cooling} could compress the muons' phase space 
	by a factor of $10^{5}$--$10^{6}$, leading to dense bunches of 
	$5\times 10^{12}$ muons at $200\mevc$.  In the simplest version of 
	ionization cooling, passage through matter degrades a muon's 
	longitudinal and transverse momentum in proportion.  An RF cavity adds 
	longitudinal momentum.  Iterating these steps cools the beam in the 
	transverse dimensions.  An important refinement uses wedge-shaped 
	degraders in a region of high dispersion, so that high-momentum muons 
	see more material than low-momentum muons.  By this device one can 
	cool the beam in both longitudinal and transverse dimensions.
	
	\item  A muon acceleration system to raise the captured muons quickly 
	to the desired energy.  An example presented at the Workshop consists 
	of a series of three recirculating linear accelerators (RLAs), whose 
	properties are summarized in Table 1.  The muons are 
	raised in steps from 1 to $10\gev$, from 10 to $70\gev$, and from 70 
	to $250\gev$.  Notice that the number of turns in each RLA is quite 
	small: 9, 11, and 12.  The decay losses in the RLAs, while not 
	crippling, are noticeable.  From the acceleration system, the muons 
	would be passed to a collider ring of quite modest dimensions.
\end{itemize}
We see that while the front end of a muon collider is small, it is 
also complex.  The important questions to answer are whether the 
construction and operation of such a device is feasible, and whether 
the size or the complexity is decisive in determining its cost.

\begin{table}[tb]
\label{rla_tab}
\caption{Recirculating linear accelerator parameters.}
\begin{tabular}{lccc}
 & RLA~1 & RLA~2 & RLA~3 \\
\hline
Input energy [GeV] & 1.0 & 9.6 & 70 \\
Output energy [GeV] & 9.6 & 70 & 250 \\
Turns & 9 & 11 & 12 \\
Linac length [m] & 100 & 300 & 533.3 \\
Arc length [m] & 30 & 175 & 520 \\
Bunch length [ps] & 158 & 43 & 19 \\
Revolution time [$\mu$s] & 0.9 & 3.1 & 7.0 \\
Decay losses & 9.0\% & 5.2\% & 2.4\% \\
Initial muons per bunch &$5 \times 10^{12}$ & $4.6 \times 10^{12}$ &
$4.3 \times 10^{12}$ \\
$\mu^+$ bunches per sec & 15& 15& 15 \\
\end{tabular}
\end{table}
\section*{A Higgs Factory}
The important possibility that a \mumu\ can operate as a Higgs factory 
has been studied extensively \cite{higgs} and received considerable 
attention at the Workshop \cite{higgz}.  If the Higgs boson is light 
($M_{H}\ltap 2M_{W}$), and therefore narrow, then the muon's large 
mass makes it thinkable that the reactions
\begin{displaymath}
	\mu^{+}\mu^{-} \rightarrow H \rightarrow b\bar{b}\hbox{ and other 
	modes}
\end{displaymath}
will occur with a large rate that will enable a comprehensive study 
of the properties of the Higgs boson.  We assume that a light Higgs 
boson has been found, and that its mass has been determined with an 
uncertainty of $\pm(100\,\hbox{-}\,200)\mevcc$ \cite{gunion}.  Then 
suppose that an optimized machine is built with $\sqrt{s} = M_{H}$.

The muon's mass confers another important instrumental advantage: the 
momentum spread of a muon collider is naturally small, and can be made 
extraordinarily small.  The Higgs factory can operate in two modes:
\begin{itemize}
	\item  modest luminosity ($0.05\fb^{-1}/\hbox{year}$) and high 
	momentum resolution ($\sigma_{p}/p = 3 \times 10^{-5}$);

	\item  standard luminosity ($0.6\fb^{-1}/\hbox{year}$) and momentum 
	resolution ($\sigma_{p}/p = 10^{-3}$).
\end{itemize}
At high resolution, the spread in c.m.\ energy is comparable to the 
natural width of a light Higgs boson: $\sigma_{\sqrt{s}} \approx 
\hbox{a few}\mev \approx \Gamma(H\rightarrow\hbox{ all})$.  At normal 
resolution, $\sigma_{\sqrt{s}} \gg \Gamma(H\rightarrow\hbox{ all})$.  

Parameters of the Higgs factories are given in Table 2, along with 
those of other candidates for a First Muon Collider \cite{ank}.  It is 
worth remarking that the Higgs factory would be small, with a 
circumference of just 380 meters, and that the number of turns a muon 
makes in one lifetime is 820.

\begin{table}[b!]
\label{fmc_tab}
\caption{Parameters considered at the Fermilab workshop for 
narrow-band and broad-band Higgs factories, a LEP2 
equivalent, a top factory, and a $\cfrac{1}{2}$-TeV FMC.}
\begin{tabular}{lccccc}
$\sqrt{s}$ [GeV] & 100 & 100 & 200 & 350 & 500 \\
\hline
Momentum spread, $\sigma_p/p$ &$3\times 10^{-5}$ &$1 \times 10^{-3}$&$1\times 10^{-3}$
&$1\times 10^{-3}$&$1\times 10^{-3}$ \\
Muons per bunch &$3\times 10^{12}$ &$3 \times 10^{12}$&$2\times 10^{12}$
&$2\times 10^{12}$&$2\times 10^{12}$ \\
Number of bunches& 1 & 1 & 2 & 2 & 2 \\
Repetition rate [Hz] & 15 & 15 & 15 & 15 & 15\\
$\epsilon_\perp$ [mm$\cdot$mr] &$297\pi$&$85\pi$&$67\pi$&$56\pi$&$50\pi$\\
Circumference [m] &380&380&700&864&1000\\
$f_{\mathrm{rev}}$ [Hz] &$7.9 \times 10^5$&$7.9 \times 10^5$&$4.3 \times 10^5$
&$3.5 \times 10^5$&$3.0 \times 10^5$ \\
Turns per lifetime & 820 & 820 & 890 & 1260 & 1560 \\
$\beta^\star$ [cm] & 13 & 4 & 3 & 2.6 & 2.3 \\
$\sigma_z$ [cm] & 13 & 4 & 3 & 2.6 & 2.3 \\
$\sigma_r$ [$\mu$m] & 286 & 85 & 47 & 30 & 22\\
${\mathcal{L}}_{\mathrm{peak}}\flux$ &$6 \times 10^{32}$&$7 \times 10^{33}$
&$6 \times 10^{33}$&$1 \times 10^{34}$&$2 \times 10^{34}$\\
${\mathcal{L}}_{\mathrm{av}}\flux$ &$5 \times 10^{30}$&$6 \times 10^{31}$
&$1 \times 10^{32}$&$3 \times 10^{32}$&$7 \times 10^{32}$\\
\end{tabular}
\end{table}

The first order of business is to run in the high-resolution mode to 
determine the Higgs-boson mass with exquisite precision.  The 
procedure contemplated is to scan a large number of points (determined 
by $2\Delta M_{H}/\sigma_{\sqrt{s}} \approx 100$), each with enough 
integrated luminosity to establish a three-standard-deviation 
excess.  If each point requires an integrated luminosity of 
$0.0015\fb^{-1}$, then the scan requires $100 \times 0.0015\fb^{-1} = 
0.15\fb^{-1}$, about three nominal years of running.  The reward is 
that, after the scan, the Higgs-boson mass will be known with an 
uncertainty of $\Delta M_{H} \approx \sigma_{\sqrt{s}} 
\approx 2\mevcc$, which is quite stunning.

Extended running in the form of a three-point scan of the Higgs-boson 
line at $\sqrt{s} = M_{H}, M_{H}\pm \sigma_{\sqrt{s}}$ would then 
make possible an unparalleled exploration of Higgs-boson properties.  
With an integrated luminosity of $0.4\fb^{-1}$ one may contemplate 
precisions of $\Delta M_{H}  \approx  0.1\mevcc$, 
$\Delta \Gamma_{H}  \approx  0.5\mev \approx \cfrac{1}{6}\Gamma_{H}$,
$\Delta(\sigma\cdot B(H\rightarrow b\bar{b}))  \approx  3\%$, and
$\Delta(\sigma\cdot B(H\rightarrow WW^{\star}))  \approx 15\%$. 

These are impressive measurements indeed.  The width of the putative 
Higgs boson is an important discriminant for supersymmetry, for it can 
range from the standard-model value to considerably larger values.  
Within the minimal supersymmetric extension of the standard model 
(MSSM), the ratio of the $b\bar{b}$ and $WW^{\star}$ yields is 
essentially determined by $M_{A}$, the mass of the \textsl{CP}-odd Higgs 
boson.  In the decoupling limit, $M_{A}\rightarrow \infty$, the MSSM 
reproduces the standard-model ratio.  Deviations indicate that $A$ is 
light.  In the most optimistic scenario, this measurement could 
determine $M_{A}$ well enough to guide the development of a 
second (\textsl{CP}-odd) Higgs factory using the reaction 
$\mu^{+}\mu^{-}\rightarrow A$.

Again, these remarkable measurements exact a high price.  At the 
Workshop luminosity of $0.05\fb^{-1}/\hbox{year}$, it takes 8 years to 
accumulate $0.40\fb^{-1}$ \textit{after the scan} to determine $M_{H}$ within 
machine resolution.  It is plain that this program becomes 
considerably more compelling if the Higgs-factory luminosity can be 
raised by a factor of 2 or 3---or more!

These projections are based on theorists' simulations; more attention 
is needed to experimental realities.  Precision measurements at LEP 
and SLC have benefitted from excellent determinations of the 
luminosity ${\mathcal L}$, the beam energy, and the lepton 
polarization.  For a muon collider, it has been shown that the muon 
spin tune $\gamma(g_{\mu}-2)/2$ offers a means of determining the beam 
energy to a few parts per million and the lepton polarization in real 
time \cite{monitor}.  Exploiting the fact that, for a muon collider 
ring with $\sqrt{s} \approx M_{Z}$ the muon's spin approximately flips 
from turn to turn, one measures the decay-electron energy spectrum as 
a function of turn number.  The frequency of the spin oscillations 
yields the Lorentz factor $\gamma$, and hence the beam energy, while 
the amplitude of the modulations in the energy spectrum is a measure 
of the beam polarization.

It is less clear how to make a precision determination of the 
luminosity.  An analogue of the standard $e^{+}e^{-}$ method of 
small-angle Bhabha monitors seems ruled out by the high flux of decay 
electrons.  Indeed, the first-pass concepts for muon collider 
detectors do not instrument a cone of $\pm(10\,\hbox{-}\,20)^{\circ}$ 
around the beam line \cite{sgback}.  For now we will assume that 
$\delta {\mathcal L}/{\mathcal L} = 10^{-3}$, but it is an important 
exercise to develop robust schemes for making this measurement.

Let us note finally that the flux of decay electrons challenges the 
operation of silicon detectors close to the interaction point 
\cite{sili}.

\section*{Other Options for the FMC}
Several other candidates for the First Muon Collider have been studied 
at this Workshop.  In order of increasing energy, they are a $Z$ 
factory, machines to explore the $W$-pair and top-pair thresholds, and 
a continuum machine operating at $\sqrt{s}=500\gev$.  The parameters 
assumed for these machines are displayed in Table 2.  It is worth 
noting that the average luminosities considered at the Workshop are 
about an order of magnitude smaller than those projected for 
$e^{+}e^{-}$ linear colliders \cite{eelum}.  Unless there are 
compensating advantages for a \mumu---the superior beam energy 
resolution, for example---the luminosity that can be 
achieved will be decisive.

A very-high-luminosity $Z$ factory, say twenty times the luminosity of 
LEP, would be a superb device for $B$ physics.  There is also 
unfinished business in the precision measurement of electroweak 
observables, particularly in light of the discrepancy between the 
value of the weak mixing parameter $\sin^{2}\theta_{W}$ inferred from 
the SLD measurement of $A_{LR}$ and the value determined from a host 
of measurements at LEP.  Alain Blondel \cite{alain} emphasized the 
desirability of controlling independently the polarizations of 
$\mu^{+}$ and $\mu^{-}$ for refining our understanding of 
$\sin^{2}\theta_{W}$.  Apart from the challenge of attaining adequate 
luminosity, an open issue for precision electroweak measurements in 
a \mumu\ is how to monitor the luminosity to high precision. 

Although a \mumu\ operating at $W^{+}W^{-}$ threshold could make 
impressive measurements of the $W$-boson mass, with $\delta M_{W} 
\approx 20\mev$ in $10\fb^{-1}$ \cite{wwmsb}, it is hard to imagine 
that this will be an important goal in the year 2010.  Experiments at 
LEP2 and the Tevatron Collider may soon give us a world average 
uncertainty approaching $50\mev$, and future running at the LEP2, the 
Tevatron, and the LHC will push the precision further.

It is possible that extensive measurements near top threshold could 
hold greater interest \cite{mikeb,tops}.  In principle, such 
measurements might yield extraordinarily precise measurements of the 
top-quark mass $m_{t}$, and give information on the strong coupling 
constant $\alpha_{s}$ and the Higgs-$t\bar{t}$ coupling $\zeta_{t}$.  
For those studies, the superb momentum spread of a \mumu---about an 
order of magnitude better that the momentum spread of a linear 
collider---could be a winning advantage.  I have to say that I am not 
convinced that the advertised determinations of $m_{t}$, $\alpha_{s}$, 
and $\zeta_{t}$ are actually attainable.  I fear that the statement 
that the ambiguity in defining $m_{t}$ is no larger than 
$\pm\Lambda_{\mathrm{QCD}}$ may be too glib.  I am also concerned that 
the theoretical link between the shape of the $t\bar{t}$ excitation 
curve and $m_{t}$, $\alpha_{s}$, and $\zeta_{t}$ is more ambiguous 
than has generally been assumed \cite{hoang}.  It is important to look 
critically at these questions as we assess the capabilities of both a 
\mumu\ and a linear collider.

Let us now look briefly at some physics prospects of a 500-GeV \mumu.  
There are rich possibilities for detailed study of the spectrum and 
properties of superpartners.  Strategies for constraining the (many) 
parameters of supersymmetric models in linear colliders have been 
documented extensively.  For the most part, the case for the study of 
supersymmetry in a \mumu\ is quite parallel to that for a linear 
collider \cite{barger,susy}.  (We have already noted the unique possibility to form the 
Higgs bosons in the $s$-channel reactions $\mu^{+}\mu^{-} \rightarrow 
h,H,A$.)  Linear colliders and \mumu s have different possibilities 
for exploiting beam polarization; how best to use polarization in a 
muon collider is a good issue for further study.
In specific cases considered at the Workshop, luminosity 
appeared to be a concern.  This was especially the case for the 
discovery and study of sleptons.  Since hadron colliders are not well 
suited to the search for sleptons, it is important that a lepton 
collider excel in slepton physics. 

If evidence for new strong dynamics represented by light-scale 
technicolor is found elsewhere, a \mumu\ will also have very 
significant capabilities for following up that discovery 
\cite{techni,weeken,bess}.  Technivector mesons with masses in the 
range $200\,\hbox{-}\,400\gevcc$ would be produced copiously even at a 
luminosity of $10^{32}\lum$ \cite{worm}.  A linear collider would offer similar 
possibilities, within the limitations of its $\sim3\%$ beam energy 
resolution.  It was recognized at this Workshop that a \mumu\ could be 
an impressive technipion factory, forming $\mu^{+}\mu^{-}\rightarrow 
\pi^{0}_{\mathrm{T}}$ at an appreciable rate \cite{techni}.  The rate for 
$e^{+}e^{-}\rightarrow \pi^{0}_{\mathrm{T}}$ is, of course, negligible.

A new element in the comparison with a linear collider is the claim by 
the DESY group \cite{bjorn} that it may be possible to increase the projected 
luminosity of a 500-GeV linear collider by more than an order of 
magnitude, perhaps to $\sim 10^{35}\lum$.  We have an obligation to 
explore how physics reach depends on luminosity for $e^{+}e^{-}$ 
linear colliders and \mumu s alike.

\section*{A $\mu \lowercase{p}$ Collider?}
If an energetic muon beam is stored in proximity to a high-energy 
proton beam, it is natural to consider the possibility of bringing 
them into collision.  One concept considered at the Workshop was to 
collide a 200-GeV muon beam with the Tevatron's 1-TeV proton beam, 
with a mean luminosity of $1.3 \times 10^{33}\lum$, for an annual 
integrated luminosity of about $10\fb^{-1}$ \cite{shiltsev}.  Such a 
machine would have an impressive kinematic reach, with $\sqrt{s} 
\approx 0.9\tev$ and $Q^{2}_{\mathrm{max}} \approx 8 \times 
10^{5}\gev^{2}$.  For comparison, the $e^{\pm}p$ collider \textsc{hera} 
currently operates with 27.5-GeV electrons on 820-GeV protons, for 
$\sqrt{s}\approx 0.3\tev$ and 
$Q^{2}_{\mathrm{max}}\approx 9\times 10^{4}\gev^{2}$.  The energy of 
the proton beam will increase over the next two years to $1\tev$, 
raising the c.m.\ energy by about 10\%.  The lifetime integrated 
luminosity of \textsc{hera} is projected as $1\fb^{-1}$.

Because of the high luminosity and the large kinematic reach, physics 
at high $Q^{2}$ is potentially very rich.
In one year of operation (\ie, at $10\fb^{-1}$), the $\mu p$ collider 
would yield about a million charged-current $\mu^{-}p \rightarrow 
\nu_{\mu}+\hbox{anything}$ events with $Q^{2}> 5000\gev^{2}$.  The 
\textsc{zeus} detector at \textsc{hera} has until now recorded 326 
charged-current events in that r\'{e}gime.  The search for 
new phenomena, including leptoquarks and squarks produced in 
$R$-parity--violating interactions, would be greatly extended.

On the other hand, the study of low-$x$ collisions appears very 
difficult because of the asymmetric kinematics and the angular cutoffs 
foreseen for detectors in the muon-storage-ring setting.  A 
general question is what kind of detectors would survive the harsh 
environment of the $\mu p$ collider.

\section*{Neutrino Beams from Stored Muons}
The idea of using stored muons to produce neutrino beams of a special 
character has arisen repeatedly.  A neutrino beam derived from the 
decay
\begin{displaymath}
	\mu^{-} \rightarrow e^{-}\nu_{\mu}\bar{\nu}_{e}
\end{displaymath}
is very different from the traditional beams derived from the decays 
of pions and kaons.  The neutrino beam generated in $\mu^{-}$ decay
contains $\nu_{\mu}$ and $\bar{\nu}_{e}$, but no $\bar{\nu}_{\mu}$, 
$\nu_{e}$, $\nu_{\tau}$, or $\bar{\nu}_{\tau}$.  It is much richer in 
electron (anti)neutrinos than a traditional neutrino beam, and muon 
\textit{neutrinos} are accompanied by electron \textit{antineutrinos.}  A 
neutrino beam derived from muon decay has therefore been seen as a way 
to remedy the absence of $\nu_{e}$ and $\bar{\nu}_{e}$ beams at 
high-energy accelerators.  The idea of storing very large quantities 
of muons---about a millimole per year---adds an important new element 
to the discussion, for now we can consider muon storage rings as 
extremely intense neutrino sources.

Neutrino beams generated by the decay of $10^{20}\,\hbox{-}\,10^{21}$ 
stored muons per year would make possible investigations of an 
entirely unprecedented nature: studies of deeply inelastic scattering 
in thin targets, and neutrino-oscillation studies over a wide range of 
distance/energy and at very great distances.

In the rest frame of the decaying muon, the distribution 
of muon-type neutrinos produced in the
decays $(\mu^{-}\rightarrow e^{-}\nu_{\mu}\bar{\nu}_{e},
\mu^{+}\rightarrow e^{+}\bar{\nu}_{\mu}\nu_{e})$ is 
\begin{displaymath}
	\frac{d^{2}N_{(\nu_{\mu},\bar{\nu}_{\mu})}}{dxd\Omega} = 
	\frac{x^{2}}{2\pi}[(3-2x)\pm(1-2x)\cos\theta]\; ,
\end{displaymath}
where $\theta$ is the angle between the neutrino momentum and the muon 
spin and $x = 2E_{\nu}/m_{\mu}$ is the scaled energy carried by the 
neutrino.  The distribution favors $x=1$ with ($\nu_{\mu}$ opposite, 
$\bar{\nu}_{\mu}$ along) the muon spin direction.  The distribution of 
electron-type neutrinos produced in $\mu^{\mp}$ decay is somewhat 
softer; it is given by
\begin{displaymath}
	\frac{d^{2}N_{(\bar{\nu}_{e},\nu_{e})}}{dxd\Omega} = 
	\frac{3x^{2}}{\pi}[(1-x)\pm(1-x)\cos\theta]\; ,
\end{displaymath}
which peaks at $x=\cfrac{2}{3}$ for 
($\bar{\nu}_{e}$ along, $\nu_{e}$ opposite) the muon spin direction.  
In a neutrino beam generated by $\mu^{-}$ decay, we would study at 
the same time, and in approximately equal proportions, the 
charged-current reactions $\nu_{\mu}N \rightarrow 
\mu^{-}+\hbox{anything}$ and $\bar{\nu}_{e}N \rightarrow 
e^{+}\rightarrow\hbox{anything}$, along with the corresponding neutral-current 
reactions in a statistical mixture. 

Let us examine the capabilities of a high-energy neutrino beam for 
deeply inelastic scattering experiments.  Two variants were considered 
at the Workshop \cite{ank}.  In the first, the 533-m straight section of RLA~3, 
the final recirculating linear accelerator in the Front End, provides 
the decay region.  Muons enter RLA~3 at $70\gev$ and are accelerated 
in 12 turns to $250\gev$.  The muon energy is therefore different on 
each turn, and increasing along the linac.  The mean neutrino energy 
$\langle E_{\nu} \rangle \approx 135\gev$.  The resulting neutrino beam 
is well collimated; at 600 meters downstream, half the neutrinos lie 
within $25\cm$ of the linac axis.  In the second scheme, a 10-meter 
straight section in a 250-GeV \mumu\ ring yields neutrinos with 
$\langle E_{\nu} \rangle \approx 178\gev$ during 1560 turns.  This 
beam is even better collimated, with about half the neutrinos within 
$15\cm$ of the axis 600 meters downstream.  The neutrino flux per year is 
prodigious, about a thousand times the flux the \textsc{nutev} detector 
received in a year of running with a traditional neutrino beam.

The gigantic flux of neutrinos from a millimole of stored muons means 
that the familiar massive neutrino detectors would be inappropriate 
devices \cite{nudet}.  Thin targets, instead of extremely massive target 
calorimeters, become the order of the day.  For example, a 1-meter 
liquid hydrogen target 600 meters downstream of RLA~3 would record 
$10^{7}$ deeply inelastic events per year.  We could therefore measure 
parton distributions of the proton directly, instead of inferring them 
from measurements made on heavy (typically, iron) targets.  The high 
rates and light targets should also make it possible to extend 
measurements of the parton distributions to smaller values of 
$x_{\mathrm{Bjorken}}$ than has been possible before in neutrino 
scattering.  The neutral-current / charged-current ratio could be 
measured with tiny statistical error, making possible an indirect 
measurement of the $W$-boson mass with $\delta M_{W} = 
(20\,\hbox{-}\,50)\mevcc$.  By reconstructing $10^{5}$ charmed 
particles per year, we could make improved measurements of the 
quark-mixing matrix element $|V_{cd}|$ and significantly advance our 
knowledge of the strange quark and antiquark distributions within the 
nucleon.

There are other possibilities as well.  Polarized targets might make 
it possible to probe details of the distribution of spin within the 
proton, perhaps even to study the polarization of minority components 
like the $s$ and $\bar{s}$ sea.  And we could consider the uses of 
high-resolution silicon detectors for special studies involving heavy 
flavors.

Neutrino beams from muon decay offer dramatic new possibilities for 
the study of neutrino oscillations.  The paucity of electron 
neutrinos and antineutrinos in traditional neutrino beams is the 
reason why we have limited knowledge of $\nu_{e} \leftrightarrow 
\nu_{\tau}$ oscillations: the $\bar{\nu}_{e}$ available at reactors 
are too low in energy to permit $\tau$-lepton appearance experiments.  
That limitation would be removed with muon-decay neutrino beams.  In 
addition, the intense fluxes will permit flexible experimentation over great 
distances.

Consider a beam of $\nu_{\mu}$ and $\bar{\nu}_{e}$ produced in 
$\mu^{-}$ decay.
In a detector that can measure the charge of leptons produced in 
charged-current interactions, it will be possible to distinguish the 
expected reactions 
\begin{displaymath}
	\nu_{\mu}N \rightarrow \mu^{-}+\hbox{anything and }
	\bar{\nu}_{e}N \rightarrow e^{+}+\hbox{anything}
\end{displaymath}
from the oscillation-induced reactions
\begin{displaymath}
	(\nu_{\mu}\rightarrow\nu_{e})N \rightarrow e^{-}+\hbox{anything and }
	(\bar{\nu}_{e}\rightarrow\bar{\nu}_{\mu})N \rightarrow 
	\mu^{+}+\hbox{anything} \; .
\end{displaymath}
In addition to these appearance experiments (of a new and interesting 
kind), we can look for distortions of the charged-lepton energy 
spectra that might signal oscillations.  For beams of sufficiently 
high energy, it will also be possible to perform appearance 
experiments in search of
\begin{displaymath}
	(\nu_{\mu}\rightarrow\nu_{\tau})N \rightarrow \tau^{-}+\hbox{anything and }
	(\bar{\nu}_{e}\rightarrow\bar{\nu}_{\tau})N \rightarrow 
	\tau^{+}+\hbox{anything} \; .
\end{displaymath}

Steve Geer has made a preliminary study of the fluxes and event rates 
that could be anticipated from a muon storage ring \cite{sgnu}.  A 
rough optimization of a storage ring to maximize the neutrino flux in 
a given direction results in a ring that consists of two semicircular arcs and two 
straight sections, with all segments of equal length.  In this way, 
25\% of the muons decay while pointing at the detector.  In the 
conceptual designs under consideration, the typical length of an arc 
(hence, of a straight section) is about
\begin{displaymath}
	\ell = 75\hbox{ m} \times \left(\frac{p_{\mu}}{40\gevc}\right)\; ,
\end{displaymath}
which is short.  Accordingly, it is reasonable to consider installing 
a ring sloped at a steep angle to point to a distant detector \cite{helium}.
Some interesting possibilities are presented in Table 3.  
In the case of conventional neutrino beams from meson decay, which 
require a decay region about a kilometer long, tunneling costs 
threaten to become prohibitive for dip angles greater than a few 
degrees.

\begin{table}[b!]
	\caption{Possible sites for long-baseline neutrino experiments \mbox{using} 
	beams generated in a muon storage ring at Fermilab.}
	\begin{tabular}{lccc}
		Location & Distance [km] & Dip Angle & Heading  \\
		\hline
		Soudan Mine, Minnesota & 729 & $3^{\circ}$ & $336^{\circ}$  \\
		Gran Sasso, Italy & 7332 & $35^{\circ}$ & $50^{\circ}$  \\
		Kamioka Mine, Japan & 9263 & $47^{\circ}$ & $325^{\circ}$  \\
		\hline
	\end{tabular}
	\label{lbtab}
\end{table}

Not only are the dimensions (including the maximum depth) of the muon 
storage ring reasonable, the fluxes at distant detectors are 
impressively large.  Geer has estimated that a 20-GeV muon beam would 
generate a flux of a few$\times 10^{10}\;\nu/\hbox{m}^{2}/\hbox{year}$ 
at the Gran Sasso Laboratory, some 7332~km from Fermilab.  [A useful 
comparison may be that the \textsc{nutev} detector saw a flux of about 
$10^{9}\;\nu/\hbox{m}^{2}/\hbox{minute}$ in the 1997 run.]  The 
fluxes at the Soudan Mine in Minnesota would be about a hundred times 
larger, and ten times the flux planned for the \textsc{minos} experiment.

Since an important figure of merit for neutrino-oscillation searches 
is $L/E$, the ratio of path length to neutrino energy, it may be 
advantageous to keep the muon energy low.  For 20-GeV muons, about 100 
charged-current events would occur per kiloton per year in the Gran 
Sasso.  Both the fluxes and the rates rise with muon-beam energy, but 
there is a price to pay in $L/E$.

The properties of neutrino beams produced in the decay of large 
numbers of muons are altogether very remarkable.  The possibilities 
for experiments are quite astounding.  We need to ask what a 
plausible experimental program might be, and whether the experiments 
are merely amazing, or truly interesting.  We also need to ask the 
important practical question: can this really be done?

\section*{Summary Remarks}
We do not yet know whether a \mumu\ will be a practical tool for 
particle physics, but the animated discussions at this Workshop and 
the diversity of ideas reported in this volume are evidence that the 
prospect of a \mumu\ gives us much to think about.  Some of the 
possibilities I have discussed in this short summary, as well as 
others to be found elsewhere in these Proceedings, represent 
opportunities that are both unique and remarkable.  This has been an 
unusually stimulating workshop, for the novelty and reach of the 
ideas we have discussed.  An important conclusion is that the 
campaign to explore the feasibility and utility of a \mumu\ is 
serious---and fun.

The original motivation for the \mumu\ remains the central goal: a 
practical lepton collider with multi-TeV beams.

I would like to conclude with a few general observations inspired by 
what we have heard during the Workshop.
\begin{itemize}
	\item The various machines discussed as the First Muon Collider (which 
	some have called the Next Lepton Collider) all are luminosity poor.  
	The interesting---and unique---program that has been outlined for a 
	Higgs factory would be a far more compelling prospect if it could be 
	carried out over a few years, rather than a decade.
	
	\item A program that includes many collider rings dedicated to 
	specific studies: a Higgs factory, a top factory, a $\cfrac{1}{2}$-TeV 
	collider, etc., appears very rich.  We have to keep in mind the 
	realities of the muon economy: not all elements of a multiring complex 
	will operate at once.  That means that different kinds of experiments 
	will necessarily be sequential or interleaved.  We cannot ignore 
	\textit{when} experiments might be done when we try to assess the 
	impact they will have on physics.
	
	\item  Even modest polarization can be highly useful, especially if 
	it can be controlled flexibly, and separately for $\mu^{+}$ and 
	$\mu^{-}$.  It is an advantage if polarization can be reversed on 
	demand.

	\item  Single-muon-ring devices do not seem to lack intensity.  The 
	capabilities of the intense neutrino beams produced in the decays of 
	stored leptons appear very well matched to the demands of the physics.
\end{itemize}
It is important for us to learn whether a \mumu\ should be part of 
our future.  I see four important short-term goals. \P\ Determine 
the overall feasibility of the muon-collider idea, with the goal of a 
high-performance, low-cost lepton collider that reaches several TeV. 
\P\ Learn whether it is possible to build a \mumu\ as a Higgs factory, 
with adequate luminosity to carry out the initial survey in only a few 
years and growth potential to make it worthwhile to exploit Higgs 
physics for a decade.  \P\ Make serious designs of muon storage rings 
as neutrino sources and investigate their potential for transforming 
neutrino physics.  It is possible that this approach to neutrino 
physics might make sense even before we know whether a muon collider 
is viable. \P\ Develop realistic conceptual designs for muon-collider 
detectors, paying careful attention to the challenges of the 
experimental environment, especially for heavy-flavor tagging.  
Explore adventurous designs for neutrino detectors that would take 
advantage of the unique character of muon-produced neutrino beams.

In assessing all the possibilities for muon-collider physics and for 
the physics opportunities that arise from the front end of a muon 
collider, we must judge as carefully as we can what will be the scientific 
impact of experiments we could carry out using these adventurous new 
devices.  The idea of a \mumu\ is bold indeed; it calls for bold 
experiments that can change the way we think.
	
\section*{Acknowledgements}
It is a pleasure to thank Steve Geer and Rajendran Raja for the 
stimulating and pleasant atmosphere of the workshop.  I am grateful 
to the workshop staff for providing me with instantaneous copies 
of transparencies throughout the week.  
I thank the working-group convenors for advice and assistance in the 
preparation of this talk.


\begin{references}
\bibitem{helium} This is not to say that it is entirely trivial to 
operate cryogenic systems for the superconducting magnets of the muon 
storage ring with a large pressure head from the top of the ring to 
the bottom.
\end{references}
\end{document}